\documentclass[aps,pra,reprint,twocolumn,superscriptaddress,showpacs,showkeys,bibliography]{revtex4-1}
\usepackage{amsmath,amssymb,amstext}
\usepackage[usenames,dvipsnames]{color}
\usepackage{graphicx}
\usepackage[english]{babel}
\usepackage{wasysym}
\usepackage{bm,bbold,braket}
\usepackage{natbib}
\usepackage{txfonts, comment,stmaryrd}
\usepackage{dcolumn}
\usepackage{graphicx}
\usepackage{subfigure}
\usepackage{float}
\usepackage[dvipsnames]{xcolor}
\usepackage[colorlinks,bookmarks=false,citecolor=blue,linkcolor=red,urlcolor=blue]{hyperref}

\begin{document}
\title{Density engineering via inter-condensate dipole-dipole interactions}

\author{Pranay Nayak}
     \thanks{These authors contributed equally to this work.}
     \affiliation{Department of Physics, Indian Institute of Science Education and Research, Pune 411 008, Maharashtra, India}   
     \affiliation{Department of Physics, Stockholm University, SE-10691 Stockholm, Sweden}
      \author{Ratheejit Ghosh}
           \thanks{These authors contributed equally to this work.}
      \affiliation{Department of Physics, Indian Institute of Science Education and Research, Pune 411 008, Maharashtra, India}       
\author{Rejish Nath}
      \affiliation{Department of Physics, Indian Institute of Science Education and Research, Pune 411 008, Maharashtra, India}
\date{\today}

\begin{abstract}
We study the effect of inter-condensate dipole-dipole interactions in a setup consisting of physically disconnected, single-species dipolar Bose-Einstein condensates. In particular, making use of the long-range and anisotropic nature of dipole-dipole interactions, we show that the density of a {\em target} dipolar Bose-Einstein condensate can be axially confined and engineered using a trapped {\em control} dipolar condensate. Increasing the number of control condensates leads to exotic ground state structures, including periodic patterns in the target condensate. The latter leads to a structural transition between single and double-peaked structures with coherence between the peaks controlled via the separation between the control condensates.\end{abstract}
\maketitle

\section{Introduction}
%
The anisotropic and long-range nature of the dipole-dipole interactions (DDIs) led to a rich physics in dipolar quantum gases \cite{bar08,lah09,bar12,def23}. In particular, self-confined multi-dimensional bright solitons \cite{ped05, tik08,nat09, rag15, mis16} and ground states with density patterns \cite{gor00,dut07,ron07,wil08,luh10, cho23} remain a focus of study in dipolar Bose-Einstein condensates (DBECs). The discovery of dipolar quantum droplets \cite{kad16,fer16, cho16, sch16,bai16,sch22}, resulting from the interplay between contact and dipolar interactions, together with the quantum stabilization \cite{wac16,mis20} led to a lot of exciting developments in the recent past. The most remarkable among them is the observation of periodic array of droplets, both incoherent and droplet supersolids \cite{soh21,cho19,tan19,bot19,guo19,tanz19,ilz21,nat19,her21,tan21, nor21,bla22}. The coherence in a droplet supersolid can be probed via excitations \cite{guo19, tanz19,nat19,muk23} and the quantum moment of inertia \cite{tan21}. More exotic and multi-dimensional supersolids are also predicted to exist in BECs \cite{hen10,zhe15,bai18,hert21,pol21,you22,sch22,bla22,gho22,muk23b,mis23}.

Because of the long-range nature, physically disconnected dipolar quantum systems can exhibit collective phenomena \cite{kol08,hua09, lut10,bar12,kna12}. In condensates, they include dipolar drag \cite{mat11,gal14}, hybrid excitations \cite{kla09, hua10,mal13}, soliton complexes \cite{nat07,kob09,lak12,lakom12,elh19,heg21}, and coupled density patterns \cite{lako12, shr23}. The inter-layer interactions can also significantly affect the stability of an atomic dipolar condensate, despite having a small atomic dipole moment (6$\mu_B$ with $\mu_B$ the Bohr magneton for a chromium atom) \cite{mul11, wil11}. Remarkably, a recent experiment using cold gas of dysprosium atoms (10$\mu_B$) demonstrated strong inter-layer dipolar effects by reducing the layer separations to 50 nm from typical lattice spacings of 500 nm \cite{lid23}, opening up new directions in the physics of dipolar gases and possibly observe various exotic bilayer phenomena \cite{wan06, wan07,tre09, arm10, pik10,hua10,mat11,huf13,saf13,ros13,mac14, cin17,bou20,ban22}. Also, the recent developments in the experiments of polar molecules with large electric dipole moments paving alternative ways to probe inter-layer effects in dipolar gases \cite{big23, nic23}.

In this paper, motivated by the recent experimental development, we analyze the effect of inter-condensate dipole-dipole interactions in a set of well-separated condensates. In particular, we show that the ground state density of a {\em target} dipolar BEC can be engineered using an array of strongly confined {\em control} dipolar BECs. The geometry of the setup is chosen to simplify the analysis, and in particular, the target BEC is assumed to be confined in a quasi-one-dimensional trap with no axial confinement. Further,  we work in a regime where the corrections to the chemical potential from the quantum fluctuations or the Lee-Huang-Yang (LHY) corrections within each condensate can be neglected \cite{edl17,pri21}. Before indulging in condensate physics, we discuss the dipolar potential experienced by a point dipole located at the center of the target condensate due to the point dipoles placed at the control BECs. Even though the properties of a pairwise dipolar potential and their effect on condensate physics are well understood, we propose novel scenarios by exploiting the long-range and anisotropic nature of the DDIs that reveal a wider possibility of engineering the effective dipolar potential in a multi-condensate environment. A distant and localized point dipole with its dipole moment along the $z$-axis induces a double-well potential on a target dipole confined on a quasi-one-dimensional tube with its axis along the $z$ direction. An array of localized point dipoles along the $z$-axis can induce a double well lattice along the parallel Q1D tube for a target dipole.

\begin{figure}
\centering
\includegraphics[width= .9\columnwidth]{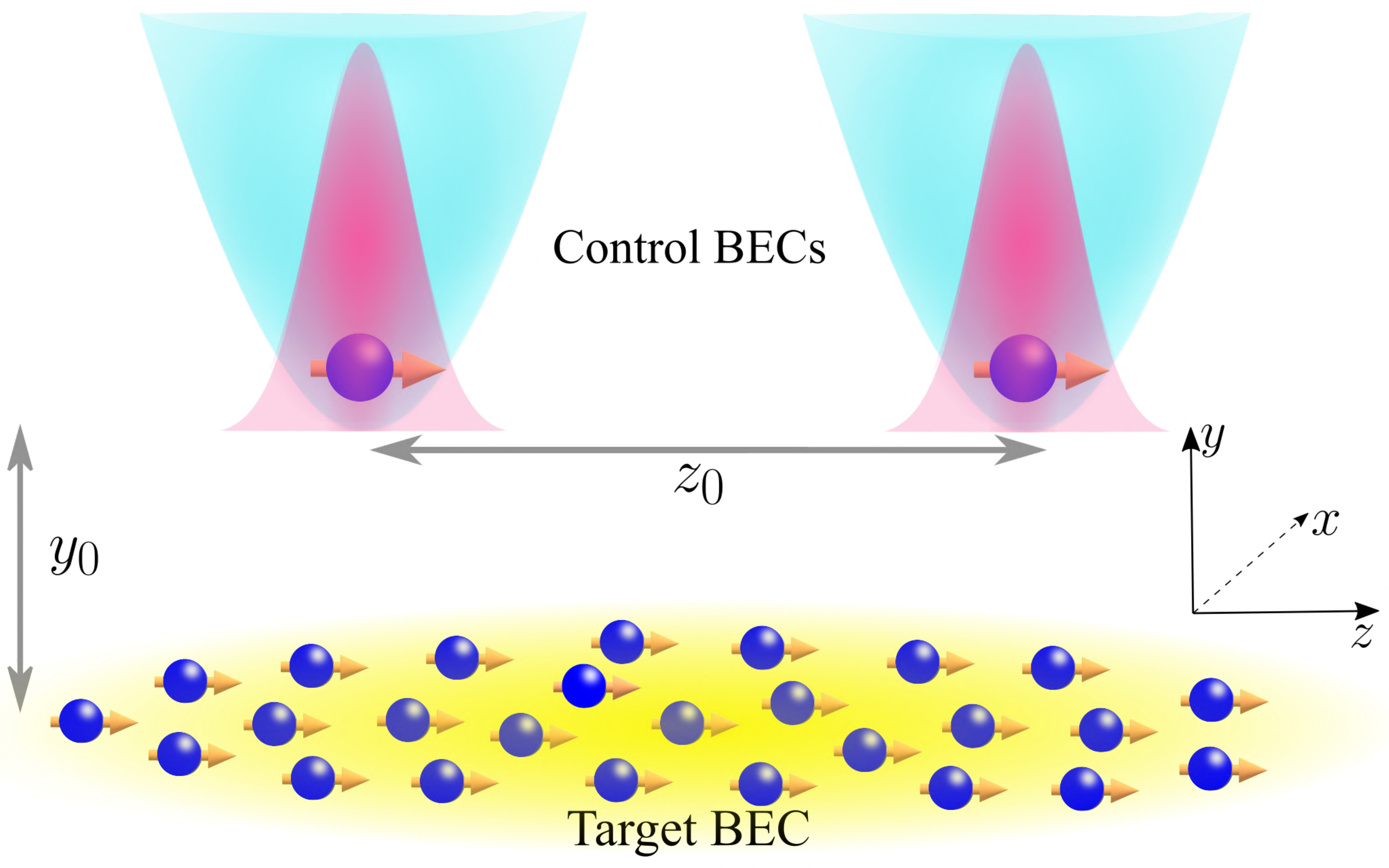}
\caption{\small{(color online). The schematic picture of the setup consists of three non-overlapping dipolar condensates. The two condensates in the top row, separated by a distance of $z_0$, are the control BECs, whereas the bottom one is the target BEC. The target BEC is separated by a distance of $y_0$ from the control ones along the $y$-axis. The control BECs are confined in all directions, whereas the target one is axially (along $z$-axis) unconfined.}}
\label{fig:0}
\end{figure}

When the point dipoles are replaced by dipolar condensates, the tightly confined control BECs act as giant dipoles and imprint double well potentials on the target condensate. The latter provides an axial confinement as well as leads to exotic density patterns in the target condensate. For instance, a single control BEC drives the target condensate into a two-peak state with the possibility of tuning the peak density by varying the trap aspect ratio of the control BEC. For the case of two control BECs,  the ground state of the target condensate undergoes a structural transition from a three-peaked to a four-peaked structure when the separation between the control BECs is increased. Interestingly, a periodic array of control BECs induces an effective periodic potential on the target BEC with a variable number of local minima, leading to different single and double-peaked periodic structures. The flow of atoms or the coherence between the density peaks can be controlled by tuning the separation between the control BECs. Since the periodic structures in target condensate are induced by the inter-condensate DDIs but not self-organized, they cannot be termed supersolid. However, they found a resemblance to immiscible double-supersolids in binary dipolar condensates in which one component supports the density modulation in the other component \cite{blan22, sha22} for which the quantum stabilization from LHY corrections is not required. The other difference is that contrary to the dipolar quantum droplets, we are at the low-density regime.

The paper is structured as follows. The setup and model are introduced in Sec.~\ref{set}. In Sec.~\ref{ic}, we discuss the axial confinement of the target dipolar BEC using a single control BEC. The density engineering using two control BECs is discussed in Sec.~\ref{eng}. The formation of periodic density patterns in the target BEC due to an array of control BECs, is discussed in Sec.~\ref{sp}. Finally, we summarize and provide an outlook in Sec.~\ref{so}.

\section{Setup and model}
\label{set}
The setup consists of $M$ spatially disconnected dipolar condensates where one is a target BEC, and the rest are control BECs. The control BECs are confined in all three directions, whereas the target one is only radially confined but not axially. The radial potential is the same for both the target and the control BECs. A schematic setup for $M=3$ is shown in Fig.~\ref{fig:0}. The two condensates in the upper row are the control BECs, and the bottom one, separated by a distance of $y_0$ in the transverse $y$-direction, is the target BEC. We consider  $N_t$ and $N_c$ number of bosons in target and control BECs, respectively. Each boson has a mass $m$ and magnetic dipole moment $d$ (results are equally valid for electric dipoles) polarized along the $z$ axis [see Fig.~\ref{fig:0}]. At very low temperatures, the system is described in the mean-field by coupled non-local Gross-Pitaevskii equations (NLGPEs),
\begin{eqnarray}
i\hbar\frac{\partial}{\partial t}\Psi_j({\bm r}, t)=\left(-\frac{\hbar^2}{2m}\nabla^2+V_j({\bm r})+gN_j|\Psi_j({\bm r}, t)|^2+\right. \nonumber \\ 
\left.\sum_{i=1}^{M}N_i\int d^3r'V_d(r-r')|\Psi_{i}({\bm r}', t)|^2\right)\Psi_j({\bm r}, t) 
\label{gpeii}
\end{eqnarray}
where $\Psi_j$ is the wavefunction of the $j$th condensate satisfying $\int |\Psi_j(r)|^2d^3r=1$, $N_j\in\{N_t, N_c\}$ and $V_j({\bm r})$ is its external potential. The external potential experienced by the control BEC is $V_j(r)=m\omega_\perp^2[x^2+(y-y_0)^2]/2+m\omega_z^2(z-z_j)^2/2$, where $\omega_\perp$ and $\omega_z$ are the trapping frequencies along the radial and axial directions and by the target BEC is $m\omega_\perp^2(x^2+y^2)/2$. The trap aspect ratio is defined as $\lambda=\omega_z/\omega_\perp$. The parameter $g=4\pi\hbar^2a_s/m$ quantifies the contact interaction strength with $a_s>0$ being the $s$-wave scattering length. The dipolar potential is $V_d(\bm{r})=g_d(1-3\cos^2\theta)/r^3$, where $g_d=\mu_0 d^2/4\pi$ with $\mu_0$ being the magnetic permeability and $\theta$ is the angle between the dipole moment and the radial vector ${\bm r}$ joining the two dipoles. We introduce the dimensionless parameters $\tilde{g}_j =gN_j/\hbar \omega_\perp l_\perp^3$ and $\tilde{g}_{dj} = g_dN_j/\hbar \omega_\perp l_\perp^3$ to quantify the interaction strengths, where $l_\perp=\sqrt{\hbar/m\omega_\perp}$. The contact interaction strength $g$ is taken sufficiently large to ensure each condensate is dynamically stable, which also rules out the self-trapping of the target BEC along the axial ($z$) axis \cite{nat07}. The ground states of the complete system are obtained by solving the coupled three-dimensional NLGPEs in Eq.~(\ref{gpeii}) via imaginary time evolution as detailed in \cite{kum15}, and our focus is particularly on the density of the target BEC. Because of the asymmetric arrangement of condensates, as shown in Fig.~\ref{fig:0}, the density of target BEC is expected to be asymmetric along the $y$-axis. But, if the system parameters and, in particular, the interaction strengths $\tilde{g}_j$ and $\tilde{g}_{dj}$ are taken such that the target BEC is in Q1D regime, the asymmetry along the $y$-axis becomes less apparent. In that case, the radial wave function can be approximated to the Gaussian ground state of the radial harmonic potential. Throughout this manuscript, we work in that regime.

\begin{figure}
\centering
\includegraphics[width= .9\columnwidth]{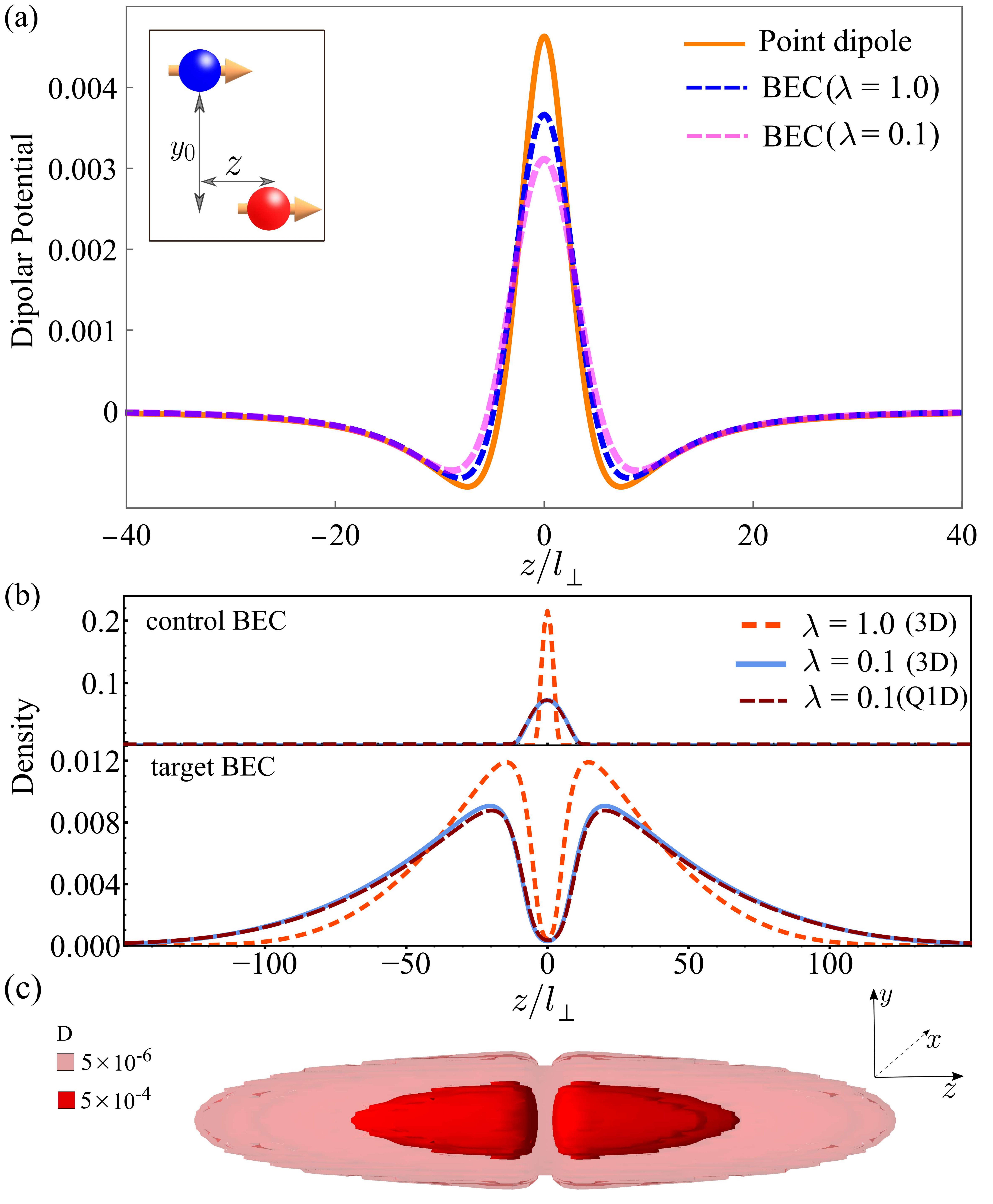}
\caption{\small{(color online). (a) The dipolar potential experienced by the target dipole due to a control dipole, $V_d^{\rm tar}(z) l_\perp^3/g_d$ as in Eq.~(\ref{pde1}) (solid line) and due to a control BEC, $\bar V_d^{\rm tar}(x=0, y=0, z) l_\perp^3/g_d$, as in Eq.~(\ref{becdp}) (dashed lines) for $y_0=6l_\perp$ and the aspect ratio $\lambda=0.1$ and 1. (b) The integrated column density  ($l_\perp\int\int dx dy|\Psi_j({ \bm r})|^2$) along the $z$-axis of control ($j=c$) and target ($j=t$) BECs for $y_0=6l_\perp$, $\tilde g_j= 210$, and $\tilde g_{dj}=50$. The interaction parameters are such that $\omega_\perp=2\pi\times 1$kHz, $N_c=N_t=9185$ ($^{52}$Cr), 1160 ($^{168}$Er) and 600 ($^{162}$Dy), and $a_s/a_0=15.2$ ($^{52}$Cr) \cite{paz14}, 66.5 ($^{168}$Er)\cite{cho19} and 131.3 ($^{162}$Dy) \cite{bot19}, where $a_0$ is the Bohr radius. Dashed lines are for $\lambda=1$, solid lines are for $\lambda=0.1$ and brown dashed lines are for Q1D results with $\lambda = 0.1$. (c) The iso-surface density plot of the target BEC for $\lambda=1$.  In the numerics, the grid extensions used are ($-x_{{\rm max}}, x_{{\rm max}}$), ($-y_{{\rm max}}, x_{{\rm max}}$) and ($-z_{{\rm max}}, z_{{\rm max}}$) with $x_{{\rm max}}=y_{{\rm max}}=25l_\perp$, and $z_{{\rm max}}=200l_\perp$, which is much larger than the size of the condensate. The cutoffs used for the dipolar potential are $R=22.5l_\perp$ and $Z=180l_\perp$. }}
\label{fig:1} 
\end{figure}
Considering our setup involves cylindrical geometries, we use a truncated dipolar potential \cite{ron06,luh10}, $V_d^{{\rm cut}}(\bm{r}) = g_{d}(1- 3 \cos^2{\theta})/r^3$ when $|z|<Z$ and $\rho=\sqrt{x^2+y^2} < R$ , and zero otherwise. Further, the DDIs in Eq.~(\ref{gpeii}) are tackled using the convolution theorem, which requires the Fourier transform of $V_d^{{\rm cut}}(\bm r)$ \cite{luh10},
\begin{align}
 \tilde{V}_d^{\rm cut}(\bm k)=4\pi g_d\left[\frac{3 \cos^2{\theta_k}-1}{3}+e^{-Zk_{\rho}}\left(\sin^2{\theta_k}\cos{k_zZ} - \frac{\sin{2\theta_k} \sin{k_{z}Z}}{2} \right)\right. \nonumber\\
\left. -\int_{R}^{\infty}\rho d\rho\int_0^{Z} dz \dfrac{\rho^2 - 2 z^2}{(\rho^2 + z^2)^{5/2}} J_{0}(k_{\rho}\rho)\cos{k_zz}\right],
\end{align}
where, $k_{\rho}= \sqrt{k_{x}^2+k_{y}^2}$, $\cos^2{\theta_k}= k_z^2/(k_{\rho}^2 + k_z^2)$, and $J_0(.)$ is the zeroth-order Bessel function. The cutoff values $Z$ and $R$ are taken such that they cover the complete system in all directions.

\subsection{Quasi-one-dimensional regime}
\label{q1d}

We can treat the whole setup in the quasi-one-dimensional (Q1D) regime when the trap aspect ratio of the control BECs is small, i.e., for $\lambda\ll 1$ and the Q1D chemical potential ($\mu_{1D}$) satisfies $\mu_{1D}\ll \hbar\omega_\perp$  \cite{lak12, lako12, heg21}. Note that, for the parameters we have taken, the target BEC is always in the Q1D regime, but for the control BECs, it depends on the value of $\lambda$. In the Q1D limit, the wavefunction of each condensate can be factorized as $\Psi_j({\bm r}, t) = \phi_{0j}(x,y)\psi_j(z, t)$, where $\phi_{0j}(x,y)$ is the Gaussian ground state of the radial harmonic potential. Employing the above factorization and integrating out the radial directions, we arrive at the coupled Q1d NLGPEs for the control BECS as,
 \begin{eqnarray}
    i\hbar\frac{\partial\psi_{j}( z, t)}{\partial t}=\Bigg[ -\frac{\hbar^2}{2m}\frac{\partial^2}{\partial z^2} + V_{j}({z}) + {g}^{1D}N_c |\psi_{j}({z}, t)|^2+ \nonumber \\
    \frac{g_d^{1D}}{3} \int \dfrac{dk_{z}}{2\pi} e^{i k_{z} z} \left(N_c\sum_{i=1}^{M-1}  \tilde n_{i}(k_z) F_{0}(k_z) +N_t \tilde n_t(k_z) F_1(k_z) \right)\Bigg] \psi_{j}(z, t),
   \label{gpeq1dc}
\end{eqnarray}
 and that for the target BEC is,
\begin{eqnarray}
    i\hbar\frac{\partial\psi_t( z, t)}{\partial t}=\Bigg[ -\frac{\hbar^2}{2m}\frac{\partial^2}{\partial z^2} + {g}^{1D}N_t |\psi_t({z}, t)|^2+  \hspace{1.6 cm}\nonumber \\
    \frac{g_d^{1D}}{3} \int \dfrac{dk_{z}}{2\pi}  e^{i k_{z} z}  \left(N_t\tilde n_t(k_z) F_{0}(k_z) + \sum_{i=1}^{M-1} N_c \tilde n_{i}(k_z) F_{1}(k_z)\right) \Bigg] \psi_t(z, t),
   \label{gpeq1dt}
\end{eqnarray}
where the sum is over all the control BECs in both the equations, $g^{1D} = g/2\pi l_{\perp}^2 $, $g_d^{1D} = g_d/2\pi l_{\perp}^2 $, $\tilde n_{i}(k_z)$ is the fourier transform of $|\psi_i(z)|^2$, and 
\begin{equation}
F_{\delta}(k_{z}) = \int d k_{x} d k_{y}\left(\frac{3 k_{z}^{2}}{k_{x}^{2}+k_{y}^{2}+k_{z}^{2}}-1\right) e^{-\frac{1}{2}\left(k_{x}^{2}+k_{y}^{2}\right) - i \delta k_{y} y_0}.
\label{eqF}
\end{equation}
The ground state solutions are then obtained by solving the coupled Q1D NLGPEs via imaginary time evolution, which are compared to the column densities obtained from the 3D calculations when $\lambda\ll 1$.

\section{Axial confinement}
\label{pd}

\begin{figure}
\centering
\includegraphics[width= 0.9\columnwidth]{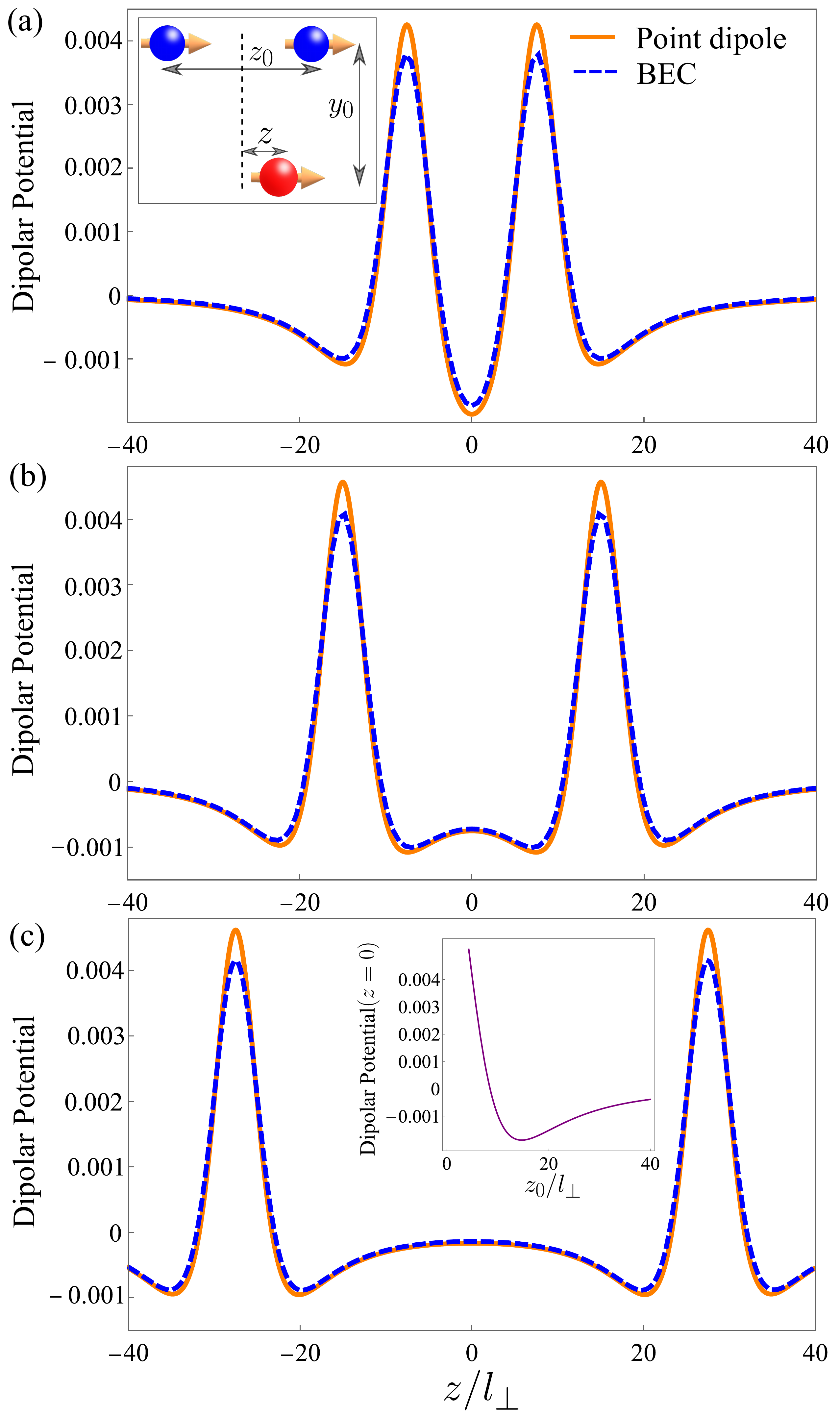}
\caption{\small{(color online). The inset of (a) shows the schematic diagram of the three-point dipole setup. The two control dipoles (blue) are shown on the top row, localized and separated by $z_0$. The dipole in the bottom (red) is the target one. The dipolar potential experienced by the target dipole due to two localized control dipoles, $V_d^{\rm tar}(z) l_\perp^3/g_d$ as in Eq.~(\ref{vdt}) (solid lines) and the same due to two control BECs for $y_0=6l_\perp$, (a) $z_0=15 l_\perp$, (b) $z_0=30l_\perp$ and (c) $z_0=55l_\perp$. The inset of (c) shows $V_d^{\rm tar}(z=0)l_\perp^3/g_d$ as a function of $z_0$, exhibiting a non-monotonous behavior. The trap aspect ratio of control BECs is $\lambda=0.33$.}}
\label{fig:2} 
\end{figure}

First, we discuss how a control BEC axially confines the target  BEC via the inter-tube DDIs. A qualitative picture can be easily obtained by considering points dipoles as follows.

\subsection{Point dipoles}
\label{scd}

The schematic diagram of a pair of point dipoles is shown in the inset of Fig.~\ref{fig:1}(a). The control dipole is localized at the coordinates $y=y_0$ and $z=0$, whereas the target dipole at $y=0$  is free to move along the axial $z$-direction. The dipolar potential experienced by the target dipole along the $z$-axis is 
\begin{equation}
V_d^{\rm tar}(z)=g_d \frac{(y_0^2-2z^2)}{(y_0^2+z^2)^{5/2}}.
\label{pde1}
\end{equation}
Since two parallel dipoles in side-by-side configuration repel each other maximally, $V_d^{\rm tar}(z)$ exhibits a finite repulsive barrier centered at $z=0$. As the separation $z$ becomes larger, the dipoles are in a head-to-tail configuration, and consequently,  the dipolar potential is attractive but decaying as $\sim -2g_d/z^3$. This attractive nature of $V_d^{\rm tar}(z)$ at large $z$ axially confines the target dipole. The change from repulsive to attractive behavior of $V_d^{\rm tar}(z)$ results in a local minimum on either side of $z=0$. Effectively, the control dipole induces a double-well potential on the target dipole, as shown (solid line) in Fig.~\ref{fig:1}(a). Note that, in general, the potential  $V_d^{\rm tar}(z)$ has radial dependence as well, which becomes irrelevant in our setup due to the strong radial confinement.

\begin{figure*}
\centering
\includegraphics[width= 1.8\columnwidth]{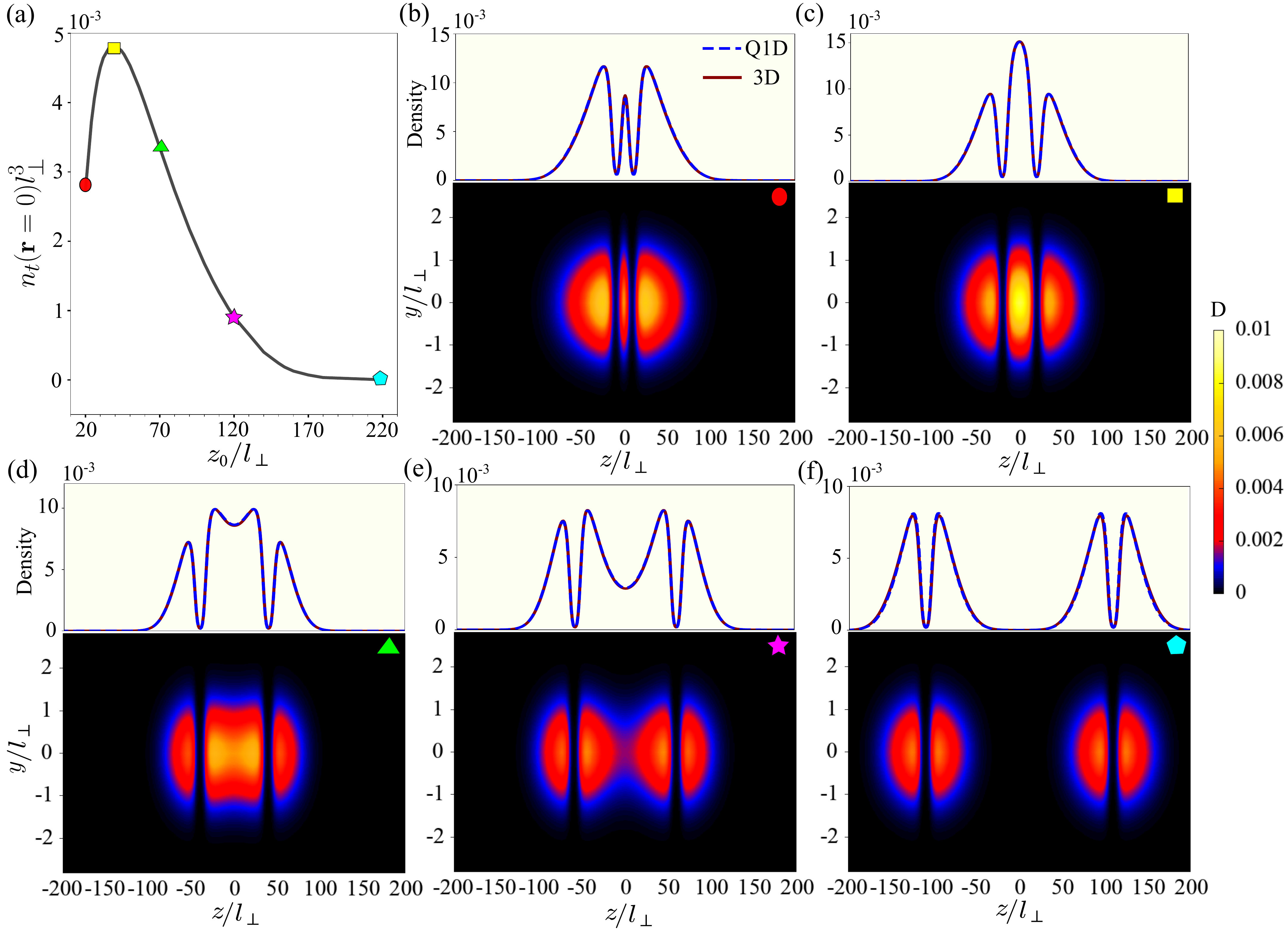}
\caption{\small{(color online). (a) The central density, $n_t({\bm r}=0)=|\Psi_t({\bm r}=0)|^2$ of the target BEC as a function of $z_0$. (b)-(f) shows the integrated ground state densities: $l_\perp\int dx dy|\Psi_t({ \bm r})|^2$ (top) and $l_\perp^2\int dx|\Psi_t({ \bm r})|^2$(bottom) of the target BEC for $z_0=20l_\perp$, $z_0=40l_\perp$, $z_0=80l_\perp$, $z_0=120l_\perp$ and $z_0=220l_\perp$, respectively and $\lambda=0.33$. (b)-(d) exhibit three peaks, whereas (e)-(f) possess four peaks. The other parameters are same as in Fig.~\ref{fig:1} and the Q1D results ($|\psi_t(z)|^2$) are in excellent agreement with the 3D calculations. D stands for the condensate density. In the numerics, the grid extensions used are ($-x_{{\rm max}}, x_{{\rm max}}$), ($-y_{{\rm max}}, x_{{\rm max}}$) and ($-z_{{\rm max}}, z_{{\rm max}}$) with $x_{{\rm max}}=y_{{\rm max}}=30l_\perp$, and $z_{{\rm max}}=300l_\perp$. The cutoffs used for the dipolar potential are $R=27 l_\perp$ and $Z=270l_\perp$. }}
\label{fig:3} 
\end{figure*}

\subsection{Dipolar condensates}
\label{ic}
At this point, we consider dipolar BECs instead of point dipoles. As the control BEC is confined in all directions, it acts as a giant dipole, and we expect the target dipole to experience a qualitatively similar potential as that of point dipoles. To verify that, we calculate the (mean-field) dipolar potential induced by a dipole distributed over the density of the control BEC on the target dipole, i.e.,
\begin{equation}
\bar V_d^{\rm tar}(\bm{r})=\int d^3r' V_d(\bm{r}-\bm{r}') n_c(\bm{r}'),
\label{becdp}
\end{equation}
where $n_c(\bm{r})=|\psi_c(\bm{r})|^2$ is the density of the control BEC. In Fig.~\ref{fig:1}(a), we show $\bar V_d^{\rm tar}(x=0, y=0, z)$ and is found to be similar to the double-well potential generated by a point dipole. The larger the value of $\lambda$, the closer the potential to the point dipole case. The actual potential on the target BEC is $N_c$ times the potential $\bar V_d^{\rm tar}(\bm{r})$.

As expected, the ground state of the target condensate is axially confined by the inter-tube DDIs. In Fig.~\ref{fig:1}(b), we show the results for two values of $\lambda$ and keep  $N_c=N_t$. The integrated column density of the control BEC is shown at the top, and that of the target BEC is shown at the bottom. The repulsive barrier in $V_d^{\rm tar}(z)$ creates a density minimum on the target BEC centered at $z=0$, leading to a two-peaked structure, as shown in Figs.~\ref{fig:1}(b) and \ref{fig:1}(c). The target BEC's axial width and peak density can be easily controlled by adjusting $\lambda$. The tighter the axial confinement of the control BEC, the larger the peak density and the lesser the axial width of the target BEC. The number of atoms required for the state-of-the-art dipolar BECs such as chromium ($^{52}$Cr) \cite{gri05, bea08}, erbium ($^{168}$Er) \cite{aik12,son23,seo23} and dysprosium ($^{162}$Dy) \cite{lum11} are provided in the caption of Fig.~\ref{fig:1} together with the required $s$-wave scattering lengths. The value of $a_s$ is such that the ratio $g_d/g$, the relevant quantity, is the same for all the species. Further, note that the Q1D (dashed thin line) results are in excellent agreement with the 3D (solid line) results, as shown in Fig.~\ref{fig:1}(b).

\section{Density Engineering of the target BEC}
\label{eng}
Increasing the number of control BECs can lead to exotic density patterns in the target BEC. To demonstrate that, first, we consider the case of two control BECs as depicted in Fig.~\ref{fig:0}. As shown below, the density patterns depend critically on the distance $z_0$ between the two control BECs. To get an intuitive picture, we again look at the dipolar potential of three-point dipoles in a geometry identical to the BEC setup shown in Fig.~\ref{fig:0}. The dipolar potential experienced by the target dipole (bottom) due to two control dipoles (top) localized at $\pm z_0/2$ is,
\begin{equation}
V_d^{\rm tar}(z)=16g_d\frac{2 y_0^2 - (z_0 - 2z)^2 }{\left[4 y_0^2 + (z_0 - 2z)^2\right]^{5/2}} + \frac{2 y_0^2 - (z_0 + 2z)^2 }{\left[4 y_0^2 + (z_0 + 2z)^2\right]^{5/2}}.
\label{vdt}
\end{equation}
In Fig.~\ref{fig:2}, we show $V_d^{\rm tar}(z)$ as a function of $z$ for three different values of $z_0$. The potential has two maxima at $\pm z_0/2$ due to the side-by-side repulsion between the control dipoles and the target dipole. A minimum is on either side of each maximum, with the total number of minima depending on $z_0$. For small values of $z_0$, there are three minima [see Fig.~\ref{fig:2}(a)]. The minimum of $V_d^{\rm tar}(z)$ at $z=0$  gets deeper as $z_0$ increases. A further increase in $z_0$, the central minimum turns into a local maximum, resulting in a pair of double-well potentials on either side of $z=0$ [see Fig.~\ref{fig:2}(b)], as expected. At large $z_0$, we get a well-separated pair of double-well potentials as shown in Fig.~\ref{fig:2}(c), arising from each control dipole. The non-monotonous behavior of the potential at $z=0$ as a function of $z_0$ is shown in the inset of Fig.~\ref{fig:2}(c). In short, the dipolar potential experienced by the target dipole can be engineered by varying the separation between the control dipoles. In particular, $V_d^{\rm tar}(z)$ exhibits three local minima for small $z_0$ and four at large $z_0$. When considering BECs, we use Eq.~(\ref{becdp}) to calculate the contribution from each control BEC and sum over them. The resulting potential for the condensates scaled by $g_d/l_\perp^3$ is also shown in Fig.~\ref{fig:2} for $\lambda=0.33$ and is almost identical to that of point dipoles.

The effect of varying the distance $z_0$ between the control BECs on the ground state of the target BEC ($n^{\rm tar}$) is shown in Fig.~\ref{fig:3}. As before, the inter-condensate DDIs axially confine the target condensate. In Fig.~\ref{fig:3}(a), we plot the central density of the target BEC as a function of $z_0$, which exhibits a non-monotonous behavior possessing a maximum, as expected from the spatial dependence of $V_d^{\rm tar}(z=0)$ shown in the inset of Fig.~\ref{fig:2}(c). For sufficiently small $z_0$, we get a three-peak structure with a sharp peak at the center [see Fig.~\ref{fig:3}(b)]. Upon increasing $z_0$, the central peak gets broader and denser [Fig.~\ref{fig:3}(c)] since the minimum of the potential at $z=0$ gets deeper. Above a particular value, the behavior changes, and an increase in $z_0$ starts to reduce the central density. As $V_d^{\rm tar}(z=0)$ becomes a local maximum, the central density lobe develops a modulation as seen in Fig.~\ref{fig:3}(d), which eventually breaks into two peaks [see Figs.~\ref{fig:3}(e) and \ref{fig:3}(f)]. Finally, at large $z_0$, a pair of semi-ellipsoid condensates is formed on either side of $z=0$, leading to a four-peaked ground state density profile. The separation between the double peak structure can be made larger by increasing $z_0$ further. Note that in each case, we have compared the column density from 3D calculations with Q1D results and found them to be in excellent agreement. As stated before, even though the presence of control BECs breaks the symmetry of the target BEC along the $y$-axis, it is not apparent in the densities shown in Fig.~\ref{fig:3} due to its Q1D nature.

\begin{figure}
\centering
\includegraphics[width= .9\columnwidth]{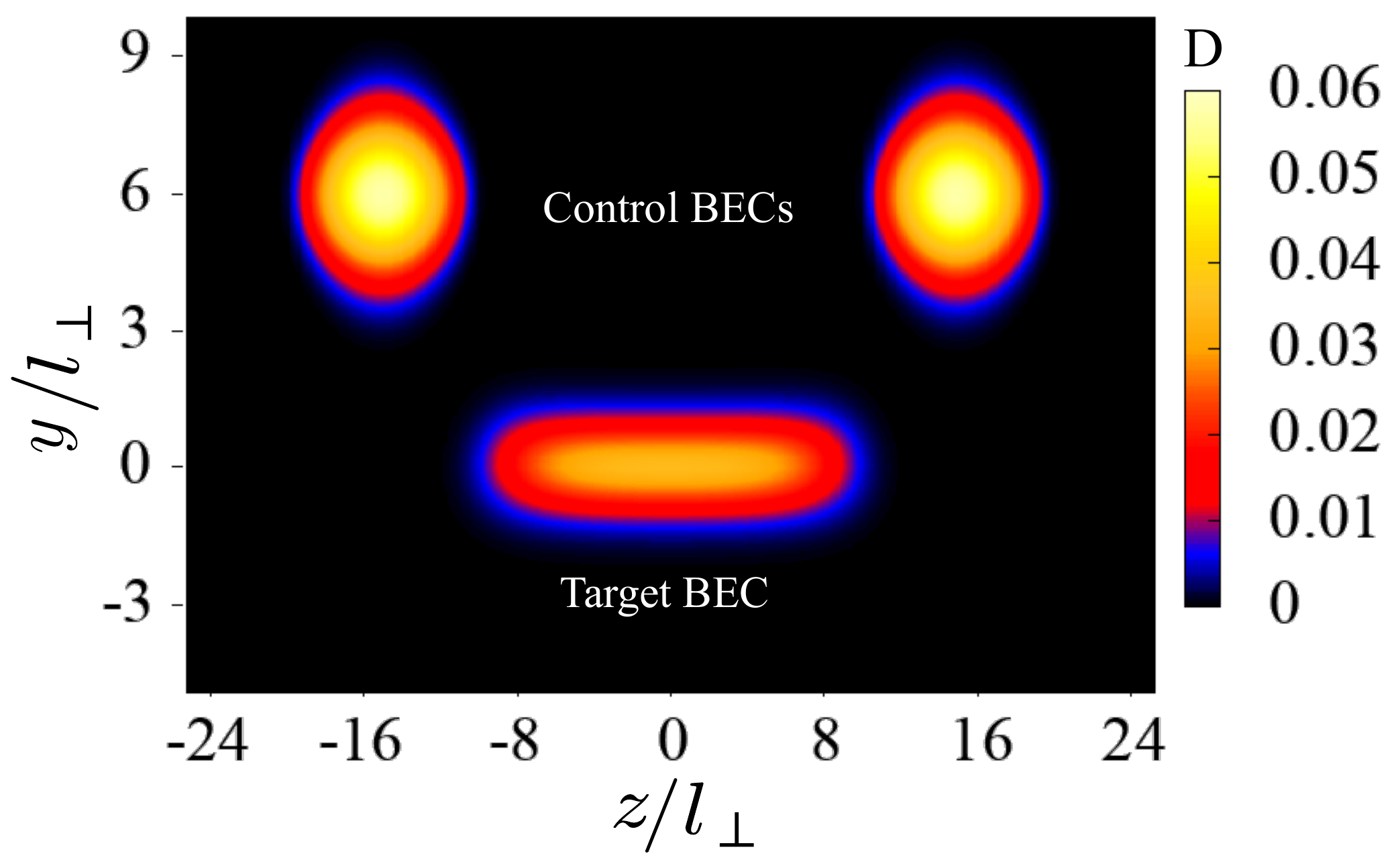}
\caption{\small{(color online). The integrated condensate density ($l_\perp^2\int dx|\Psi_j({ \bm r})|^2$) of control ($j=c$) and target ($j=t$) BECs for $\omega_\perp=2\pi\times 1$kHz, $y_0=6l_\perp$, $z_0=30 l_\perp$ and $\lambda=1$. For control BECs, $\tilde g_{dc}=200$ and $\tilde g_c=840$, and for the target BEC, $\tilde g_{dt}=10$ and $\tilde g_t=42$ and together, they correspond to $\{N_c=36740, N_t=1840\}$ for $^{52}$Cr, $\{N_c=4640, N_t=230\}$ for $^{168}$Er and $\{N_c=2400, N_t=120\}$ for $^{162}$Dy. D stands for condensate density. In the numerics, the grid extensions used are  ($-x_{{\rm max}}, x_{{\rm max}}$), ($-y_{{\rm max}}, x_{{\rm max}}$) and ($-z_{{\rm max}}, z_{{\rm max}}$) with $x_{{\rm max}}=y_{{\rm max}}=30l_\perp$, and $z_{{\rm max}}=150l_\perp$. The cutoffs used for the dipolar potential are $R=27 l_\perp$ and $Z=135l_\perp$. }}
\label{fig:4} 
\end{figure}

So far, the atom number is kept the same for target and control BECs. Reducing the number of atoms in the target BEC leads to a single ellipsoid-shaped condensate formed in the region between the control condensates, as shown in Fig.~\ref{fig:4}. In Fig.~\ref{fig:4}, we show the densities of both the control (two peaks on the top) and target (bottom) condensates.

\section{Periodic patterns}
\label{sp}

\begin{figure}
\centering
\includegraphics[width= 0.9\columnwidth]{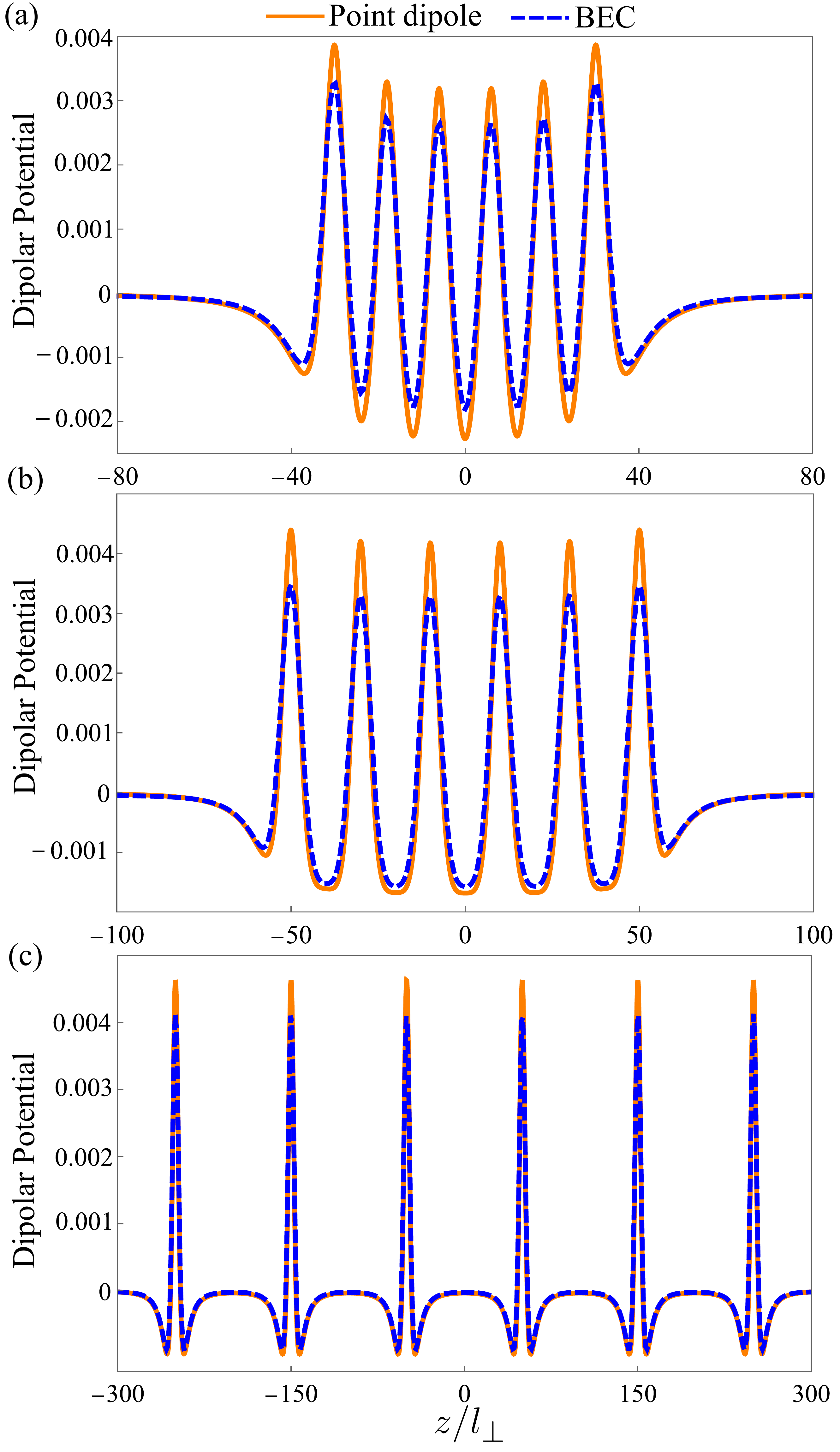}
\caption{\small{(color online). The dipolar potential $V_d^{\rm tar}(z) l_\perp^3/g_d$ experienced by the target dipole due to six localized control dipoles (solid line) for $y_0=6l_\perp$, (a) $z_0=12 l_\perp$, (b) $z_0=20 l_\perp$ and (b) $z_0=100 l_\perp$, where $z_0$ is the separation between adjacent control dipoles. The same due to control BECs for $\lambda=1$ is shown by dashed line. At large $z_0$, as in (c), the total potential can be seen as an array of double-well potentials. }}
\label{fig:5} 
\end{figure}
\begin{figure*}
\centering
\includegraphics[width= 2\columnwidth]{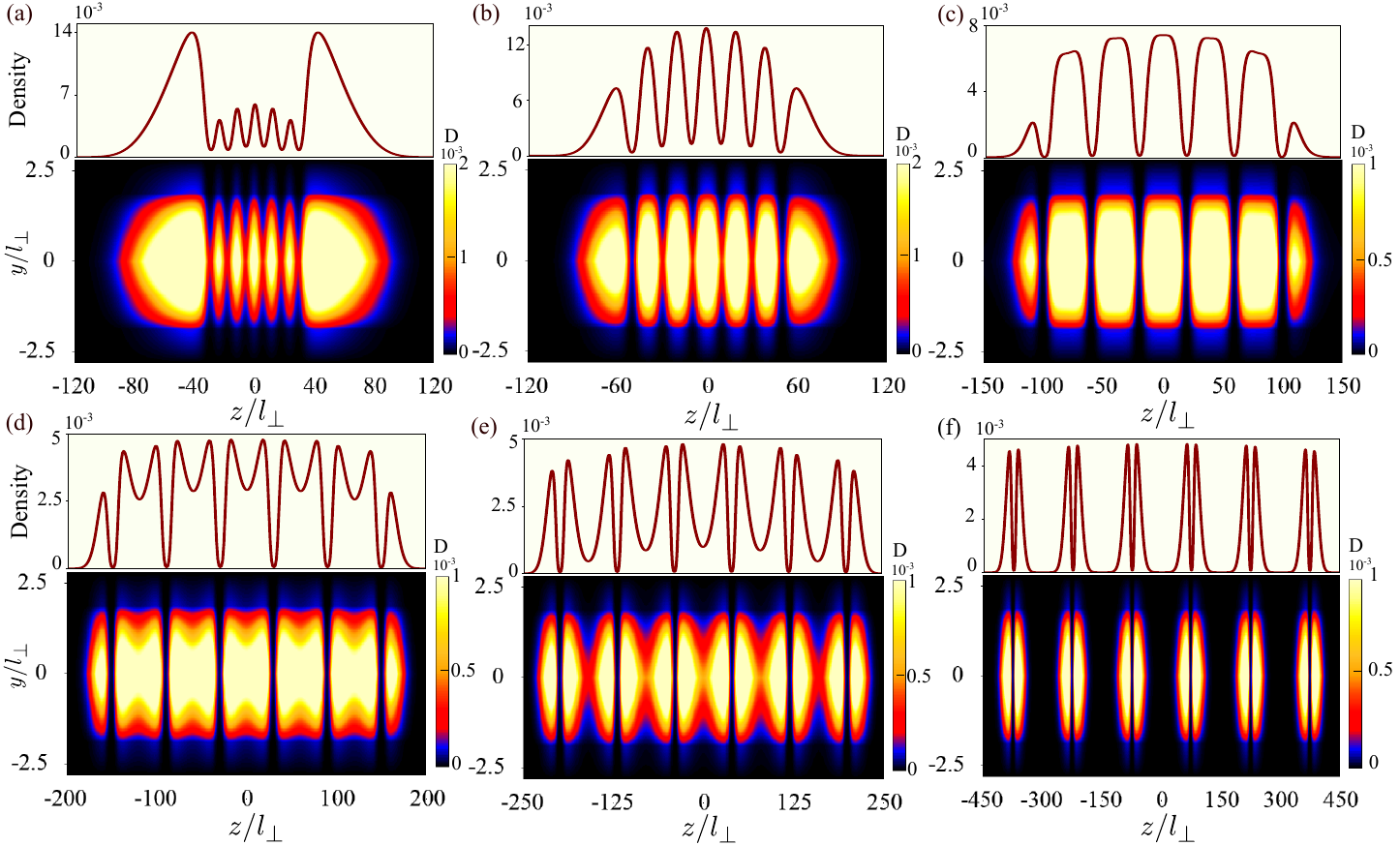}
\caption{\small{(color online). The column densities: $l_\perp\int dx dy|\Psi_t({ \bm r})|^2$ (top) and $l_\perp^2\int dx|\Psi_t({ \bm r})|^2$(bottom) of the target BEC for different values of $z_0$ in the presence of a periodic array of six control BECs for $y_0=6l_\perp$, (a) $z_0=12l_\perp$, (b) $z_0=20l_\perp$, (c) $z_0=40l_\perp$, (d) $z_0=60l_\perp$, (e) $z_0=80l_\perp$,  and (f) $z_0=150l_\perp$. The other parameters are $\lambda=1$, $\tilde g_{dt}=\tilde g_{dc}=50$ and $\tilde g_t=\tilde g_c=210$, and they are the same as for Fig.~\ref{fig:1}. D stands for condensate density. In the numerics, the grid extensions used are ($-x_{{\rm max}}, x_{{\rm max}}$), ($-y_{{\rm max}}, x_{{\rm max}}$) and ($-z_{{\rm max}}, z_{{\rm max}}$) with $x_{{\rm max}}=y_{{\rm max}}=30l_\perp$, and (a)-(b) $z_{{\rm max}}=150l_\perp$, (c) $z_{{\rm max}}=300l_\perp$, (d) $z_{{\rm max}}=350l_\perp$, (e) $z_{{\rm max}}=450l_\perp$ and (f) $z_{{\rm max}}=760l_\perp$. The cutoffs used for the dipolar potential are $R=27 l_\perp$, (a)-(b) $Z=135l_\perp$, (c) $Z=270l_\perp$, (d) $Z=315l_\perp$, (e) $Z=405l_\perp$ and (f) $Z=684l_\perp$.}}
\label{fig:6} 
\end{figure*}

Finally, we show that having an array of many control BECs can induce various periodic patterns on the ground state density of the target BEC. In Fig.~\ref{fig:5}, we show the potential experienced by the target dipole due to a periodic array of six control dipoles for different $z_0$, where $z_0$ is the separation between the adjacent control dipoles. For small $z_0$, (large enough to avoid any overlap between the control BECs) the potential has seven local minima [see Fig.~\ref{fig:5}(a) for $z_0=12l_\perp$], with inner minima being narrower than the outer ones. As $z_0$ increases, the inner minima get broader, and the outer ones get narrower  [see Fig.~\ref{fig:5}(b) for $z_0=20l_\perp$]. When the separation is more significant, each control dipole induces a double-well potential centered around its position. Hence, a maximum of twelve minima emerges in the dipolar-potential as shown in Fig.~\ref{fig:5}(c), say for $z_0=100l_\perp$. 

The numerically obtained ground state densities of the target BEC are shown in Fig.~\ref{fig:6}. For small $z_0$, we see a seven-peaked structure, with outer density lobes significantly broader and higher in density than the inner ones [see Fig.~\ref{fig:6}(a) for $z_0=12l_\perp$]. It is expected that since the inner potential minima are narrower, increasing the density will cost more energy. Increasing $z_0$ leads to a structural modification, where the inner lobes get broader and denser than those at the edges [see Figs.~\ref{fig:6}(b) and \ref{fig:6}(c) for $z_0=20l_\perp$ and $z_0=40l_\perp$]. As $z_0$ increases further, the inner density lobes develop density modulation, [see Figs.~\ref{fig:6}(d) for $z_0=60 l_\perp$] and eventually becomes double-peaked pattern at large values of $z_0$ as shown in Figs.~\ref{fig:6}(e) and \ref{fig:6}(f) for $z_0=80 l_\perp$ and $z_0=150 l_\perp$, respectively. Due to these structural modifications, as $z_0$ increases, the central density changes from a maximum to a local minimum ($z_0>40l_\perp$) and eventually vanishes.


\begin{figure}
\centering
\includegraphics[width= .9\columnwidth]{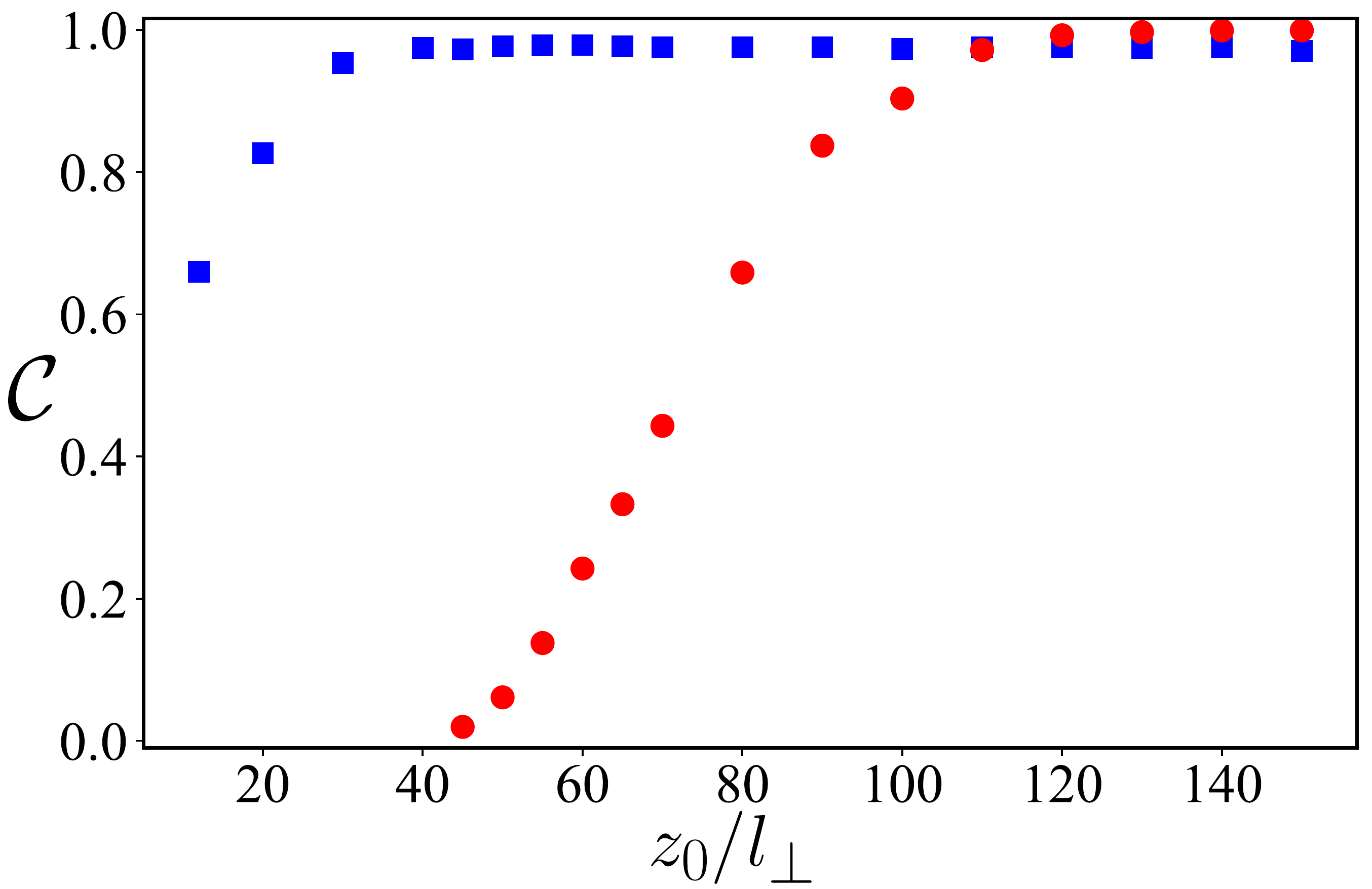}
\caption{\small{(color online). The contrast $\mathcal C$ obtained for the density patterns shown in Fig.~\ref{fig:6} as a function of $z_0$. The filled squares correspond to the contrast obtained between two central density lobes. For $z_0>40l_\perp$, each density lobe develops a density minimum, and the corresponding contrast is shown by a second branch (filled circles).  }}
\label{fig:7}
\end{figure}
At this point, we quantify the coherence between two neighboring density peaks at the center of the target condensate using the contrast \cite{blan22}, defined as  
\begin{equation}
    \mathcal C = \frac{n_{max}-n_{min}}{n_{max}+n_{min}},
\end{equation}
where $n_{max}$ is the maximum density among the peaks and $n_{min}$ is the density minimum between two peaks. For small separation between the control BECs, till $z_0=40l_\perp$, the contrast between the neighboring peaks at the central region of the condensate, shown by filled squares in Fig.~\ref{fig:7}, increases with $z_0$ and eventually attains almost a value of one. The latter indicates an incoherent array of density peaks for $z_0\sim40l_\perp$. For  $z_0>40l_\perp$, each density lobes develop a local minimum, and the contrast obtained within a given density lobe again increases with $z_0$ as shown by filled circles in Fig.~\ref{fig:7}, and eventually attaining a value of one. Again, The latter indicates an incoherent array of density peaks, but double the number of peaks compared to that for $z_0\sim 40l_\perp$.
\section{Summary and Outlook}
\label{so}

To summarize, utilizing the inter-condensate DDIs, we showed that the density of a target BEC can be axially confined and engineered using a single control BEC. When extended to multiple control BECs, exotic density patterns are formed. Each control BEC induces a double-well-like potential on the target BEC. The effective dipolar potential acting on the target BEC depends critically on the separation between the control BECs. The temperature associated with the potential minima is the order of tens of nano-kelvins for the parameters considered. It is easily tunable using separation between the target and control BECs. Interestingly, we observed a structural crossover between patterns of two periodicities when the distance between the adjacent control BECs is varied.

 The exciting aspect of our results is that once strongly confined, a dipolar BEC can affect the density of another dipolar BEC due to its long-range and anisotropic nature, even if kept far away. Even though it sounds simple, it is a non-trivial result if carefully considered. The previous studies where one condensate affects another involved multi-component BECs, which occupy the same space and are externally confined or in the self-trapping regime. Our studies also open up several perspectives. Here, we only focused on the periodic arrangement of control BECs, but an aperiodic arrangement can induce non-regular structures on the target BEC. The above studies could be extended in a 2D arrangement of control BECs. Our results, in general, indicate the possibilities of engineering the quantum state of one dipolar system using another, for instance, in a hybrid setup of polar molecules or Rydberg atoms, as they have been gaining importance recently  \cite{gut23}. 


\section{Acknowledgements}
We thank C. Mishra for the discussions during the initial stages of the work. We acknowledge National Supercomputing Mission (NSM) for providing computing resources of "PARAM Brahma" at IISER Pune, which is implemented by C-DAC and supported by the Ministry of Electronics and Information Technology (MeitY) and Department of Science and Technology (DST), Government of India. P. N. acknowledges the funding from DST India through an INSPIRE scholarship. R.N. further acknowledges DST-SERB for Swarnajayanti fellowship File No. SB/SJF/2020-21/19, and the MATRICS grant (MTR/2022/000454) from SERB, Government of India.
\bibliography{qeng.bib}
\end{document}